# Comprehensive Multimodal and Multiscale Analysis of Alzheimer's Disease in 5xFAD Mice: Optical Spectroscopies, TEM, Neuropathological, and Behavioral Investigations.

Dhruvil Solanki [1], Ishmael Apachigawo [1], Sazzad Khan [2], Santanu Maity[1], Fatemah Alharthi [1], Samia Nasim [2,3], Mohammad Alizadeh Poshtiri[1], Fnu Sweety[1], Jiangfeng Xiao [2], Mohammad Moshahid Khan [2],* and Prabhakar Pradhan [1],*

[1]*Department of Physics and Astronomy, Mississippi State University, Mississippi State, MS 39762, USA*
[2]*Department of Neurology, College of Medicine, University of Tennessee Health Science Center, Memphis, TN 38163, USA*
[3]*Department of Ophthalmology, Hamilton Eye Institute, University of Tennessee Health Science Center, Memphis, TN 38163, USA*

## Abstract

Alzheimer's disease (AD) is considered one of the leading causes of death in the United States, and there is no effective cure for it. Understanding the neuropathological mechanisms underlying AD is essential for identifying early, reliable biomarkers and developing effective therapies. In this paper, we report on a comprehensive multimodal study of AD pathology using the 5xFAD mouse model. We employed light-scattering techniques, Partial Wave Spectroscopy (PWS) and Inverse Participation Ratio (IPR), to detect nanoscale structural alterations in brain tissues, nuclear components, and mitochondria. To support the light-scattering experiments, behavior, and histopathological studies were conducted. These analyses revealed significant increases in structural heterogeneity and mass density fluctuations in the brains of 5xFAD mice compared with Non-transgenic controls. Behavioral assessment demonstrated memory impairment in 5xFAD mice, reflected by a reduced recognition index. Histopathological analysis further revealed increased amyloid beta plaques and microglia activation in the hippocampus and cortex of 5xFAD mice compared with Non-transgenic controls. An increase in structural disorder within brain tissues can be attributed to higher mass density fluctuations, likely arising from macromolecular rearrangement driven by amyloid beta aggregation and neuroinflammatory responses as the disease progresses. Our findings suggest that PWS and IPR-derived metrics provide sensitive biophysical indicators of early cellular and subcellular disruption, offering potential as quantitative biomarkers for the detection and progression of AD.

## 1. Introduction

Alzheimer's Disease (AD) is a progressive neurodegenerative disorder characterized by an impairment in cognitive function and memory, often accompanied by difficulty in doing routine activities [1]. This disease generally affects elderly people who are 65 or older, but younger people can also be affected by AD. Pathologically, AD is linked to the extracellular accumulation of amyloid beta (Aβ) plaques in the cerebral cortex. These plaques impede communication between neurons and impair function in the cerebral region. Secondly, neurofibrillary tangles composed of aberrant tau proteins within neurons and glia have been observed. Together, these pathological hallmarks contribute to synaptic dysfunction and neuronal loss [2–5]. It has been estimated that nearly 7.2 million Americans older than 65 are currently affected by AD. This number is expected to grow to 14 million by 2060 [2]. Despite extensive research, AD is rarely diagnosed at its early stages, even though pathological changes occur at subcellular levels long before clinical symptoms become apparent. Moreover, current diagnostic methods typically identify the disease only after significant neuropathology has appeared, limiting the effectiveness of therapeutic intervention. Therefore, there is an urgent need to develop a sensitive and reliable technique for early AD diagnosis. For that, we need to understand changes in brain tissue better as the disease progresses and identify more promising biomarkers. It is known that Aβ plaques predominantly accumulate in the hippocampus and cortex, which are critically involved in memory and cognition, making them essential targets for studying AD-related neuropathology [6,7]. So, we need to study those brain regions to understand neuropathology better.

As AD progresses, mass density changes at the tissue to the sub-cellular level [8]. However, these structural changes are not visible under conventional microscopy because these microscopes cannot resolve features below 200 nm. Techniques such as MRI and CT scans detect AD at a later stage, when symptoms have already manifested [9]. Therefore, to probe these changes at an early stage, we employed multiple light-scattering techniques, including Partial Wave Spectroscopy (PWS) and Inverse Participation Ratio (IPR), at the nano- to submicron scale on brain tissues from mice to quantify refractive index (RI) changes statistically. Biological cells and tissues are considered elastic scatterers of light, and variations in their refractive index provide insights into the underlying structural organization[10–12]. This technique scatters light differently at different positions due to refractive index variations, which helps us understand the formation of cellular structures. This technique has been successfully used previously in detecting different cancer stages [13,14]. We have also employed it on human AD brain tissues to differentiate different stages of AD [15]. In PWS, backscattered light intensity signals are used to quantify the refractive index variations (*dn*) and mass density fluctuations from weakly disordered media, including biological cells/tissues. Previous studies have shown that backscattered light intensity is proportional to refractive index fluctuations within tissues/cells, which are further related to mass density fluctuations at the nanometer-to-submicron level[16–18]. Using the mesoscopic theory of light, we calculated the structural disorder strength ($L_{d\text{-}PWS}$) for different cells/tissues and have exploited this parameter as a potential biomarker for PWS in disease studies [19]. Another technique, IPR, quantified as a potential biomarker by $L_{d\text{-}IPR}$ is used to probe molecular-specific structural alterations in nuclear components, including DNA and chromatin, as AD progresses. Fluorescent images of biological samples stained using DAPI are taken, and IPR analysis is performed. The methods section discusses detailed information about PWS and IPR techniques.

In this study, we perform spectroscopic analyses of brain tissues/cells from 5xFAD and Non-transgenic (Non-Tg) mice, primarily targeting the hippocampus and cortex regions, which are critically involved in memory and cognition. Using PWS and IPR techniques, we investigated structural alterations at the nanometer-to-submicron scale within these regions. The 5xFAD mouse model effectively mimics several critical features of AD, including the formation of Aβ plaques, synaptic dysfunction, mitochondrial dysfunction, progressive neuronal loss, genotoxic stress, and robust neuroinflammatory responses[20,21]. Our analyses revealed a significant increase in structural disorder in brain tissues and cells of 5xFAD mice compared to Non-Tg mice. We further inspected local structural alterations in hippocampal mitochondria by first scanning them with a Transmission Electron Microscope (TEM) and later applying IPR analysis to detect regional-level changes in AD conditions. Structural disorder was elevated in 5xFAD mitochondria than in Non-Tg mitochondria. We have further correlated these results with Aβ pathology, cognitive function, and mitochondrial DNA copy number, thereby providing an integrated multiscale perspective on AD pathology.

## 2. Materials and Methods

*Animal model*

All mouse experiments were performed in accordance with the National Institutes of Health's guidelines for the care and sse of laboratory animals, approved by our Institutional Animal Care and Use Committee. The 5xFAD transgenic mice (Jackson Laboratory; stock #006554) express a human amyloid precursor protein (APP) gene carrying three specific point mutations (I716V, V717I, and KM670/671NL), along with a human presenilin 1 (PSEN1) gene that contains two mutations (M146L and L286V). These mutations are co-inherited in the mouse line. All experimental animals were bred in-house and were heterozygous for both mutant transgenes. Genomic DNA extracted from tail biopsies was analyzed by PCR before weaning to verify the presence of the APP and PSEN1 transgenes in all mice.

### 2.1. Partial Wave Spectroscopy Experiment

#### 2.1.1. Optical Setup

A detailed picture of the setup can be found in our previous works [13,14,19]. However, a brief overview of the setup is provided here for completeness. The setup consists of a broadband white-light Xe lamp (Newport Corporation, Irvine, CA, USA) source, followed by a set of converging lenses forming a 4f system to collimate the light with Kohler illumination. The light then goes onto a right-angled prism (BRP), which directs the light to a beamsplitter, then to a 40x objective lens (NA = 0.65), where the sample is held by a XYZ motorized stage (x-y = 40 nm, z = 100 nm; Zaber Technologies, Vancouver, BC, Canada). The reflected backscattered light travels to a liquid crystal tunable filter (LCTF) (Thorlabs, Newton, NJ, USA) and then to a CCD, where the images are acquired in the visible range of 450-700 nm. With the motorized stage, the setup can measure the strength of structural disorder ($L_{d\text{-}PWS}$) at each pixel, with XY accuracy of 40 nm and Z accuracy of 100 nm.

#### 2.1.2. Measurement of Structural Disorder Strength

The fluctuating reflectance spectra $R(x,y,\lambda)$ from the sample are found to be proportional to refractive index fluctuations ($dn$) in the sample, which change as the mass density of cells/tissues in the cells/tissues changes at the nano- to sub-micron level. This has been demonstrated in our previous works [22–24]. We were able to quantify these mass density changes in terms of the structural disorder strength $L_{d\text{-}PWS}$ using the mesoscopic theory of light transport [23,25,26], assuming that the RI fluctuations are within its correlation length $Lc$, i.e., $L_d = <dn^2> \times Lc$, where $<dn^2>$ is the variance in RI. Reflectance spectra are recorded of a 3D sample as a 3D data cube with two spatial positions (x, y) and wavelength as an additional parameter, spanning the visible range from 450 to 700 nm, with reflectance averaged over the z direction.

Mesoscopic light transport theory can be applied to the transport of electrons and light in dielectric media [26,27]. Biological 3D samples are considered as several interconnected 1D weakly disordered parallel media, allowing the quasi-1D theory to be applied to them. The backscattered light $I(x,y,\lambda)$ is processed with a low-pass Butterworth filter to reduce high-frequency noise, and a low-order polynomial is applied to remove variations from the light source. Once we get the processed reflected spectra, the light is virtually divided into many 1D parallel channels by a quasi-1D approximation with a 200x200 nm (within the diffraction limit) pixel size. $<R(k)>$ rms value of reflected spectra is calculated in the range of 450-700 nm, with auto correlation length $C(\Delta k)$ [24,28,29].

$$L_{d-PWS} = \frac{B n_0^2 \langle R \rangle}{2k^2} \frac{(\Delta k)^2}{-\ln(\langle C(\Delta k) \rangle)} \qquad (1)$$

In the above equation, *B* is a calibration constant, $n_0 = 1$. $(\Delta k)^2/ln(<C(\Delta k)>)$ can be found by performing a linear fit of $ln(<C(\Delta k)>)$ vs $(\Delta k)^2$.

### 2.1.3. Sample Preparation for PWS Experiment

The University of Tennessee Health Science Center (UTHSC) provided all samples with IACUC approval. The mouse tissue samples were sectioned into 5 μm slices using a microtome, embedded in paraffin according to standard protocol, and placed on a glass slide for light-scattering experiments.

## 2.2. Inverse Participation Ratio Quantification using Confocal Microscopy and Transmission Electron Microscopy

### 2.2.1. Confocal Imaging

Confocal images targeting the cell's nucleus were captured using a Zeiss 710 confocal microscope (Carl Zeiss Microscopy, Jena, Germany), where we used Z-stack mode above and below a cell's nucleus. Pictures (4-8 micrographs) captured in the Z-stack mode were selected based on the most significant changes observed in the stack, providing the best coverage of the nuclear area. To analyze nuclear components such as DNA/chromatin, each cell type was identified based on its characteristics. Various software, such as MATLAB and ImageJ v1.54c (National Institutes of Health, Bethesda, MD, USA), were used to process the images. Similarly, tissue images from different groups were captured for further processing and calculations.

### 2.2.2. Measurement of IPR

The light intensity in the confocal images of biological samples is due to multiple scattering within cells and tissues. Scattering within cells and tissues occurs due to structural disorder, especially from components such as DNA, lipids, and proteins. Image intensity changes as the refractive index within cells and tissues varies. A change in refractive index can be linked to a change in mass density in a voxel of the cell. $I(x, y) \propto n(x, y) \propto M(x, y)$[30]. Similarly, the optical potential can be defined as

$$\varepsilon_i(x, y) = \frac{dn(x,y)}{n_0} \propto \frac{dI(x,y)}{I_0}. \tag{2}$$

In Eq.(2), $dn$ is the refractive index fluctuations, $n_0$ is the average refractive index of the medium, $dI$ is the variation in light intensity, and $I_0$ is the average light intensity coming from the cell voxel. The Hamiltonian of the system is defined using the Anderson tight-binding model [31,32],

$$H = \sum \varepsilon_i |i><i| + t\sum_{\langle ij \rangle} (|i><j| + |j><i|) \tag{3}$$

where $|i>$ and $|j>$ are the optical wave functions at the $i^{th}$ and $j^{th}$ lattice sites, $t$ is the inter-lattice hopping energy, and $\varepsilon_i$ is the optical potential at the $i^{th}$ state. Mean IPR can be derived by substituting the optical potential in the Hamiltonian for sample length $L$, such that,

$$\langle IPR \rangle_P \equiv \langle IPR \rangle_{L \times L} = \frac{1}{N}\sum_{i=1}^{N} \int_0^L \int_0^L E_i^4(x,y) dx dy \tag{4}$$

In equation 4, $E_i$ is the $i^{th}$ eigenfunction of the optical lattice of size $L \times L$, $p$ is the 2D pixel size of the confocal micrograph, and $N$ is the total number of eigenfunctions in the refractive index matrix ($N = (L/dx)^2$).

We have shown in our previous works that Mean and STD $\langle IPR \rangle_P$ are proportional to the structural disorder strength $L_d$ of similar cells, where $L_{d\text{-}IPR} = \langle \Delta n \rangle \times l_c$ [33].

$$\text{Mean}\langle IPR(L) \rangle_{L \times L} \propto L_{d-IPR} = <dn> \times l_c \tag{5}$$
$$STD \langle IPR(L) \rangle_{L \times L} \propto L_{d-IPR} = <dn> \times l_c \tag{6}$$

*2.2.3. Sample Preparation for IPR Experiment*

A fluorescent dye, DAPI, was used to stain nuclear structures, such as DNA and chromatin, allowing them to be visualized under fluorescence microscopy and helping us differentiate nuclei from the rest of the cell. The following steps were taken to stain the samples with dye. Firstly, the slides were cleansed in Phosphate Buffered Saline (PBS) with 2-4% paraformaldehyde for 5 minutes. This step was repeated 3 times before moving further. Later, the slides were stained with blue dye to label nuclear structures, such as DNA, using Prolong Diamond antifade mountant containing DAPI.

*2.2.4. TEM Imaging of Mitochondria*

Transmission electron microscopy was performed using a JEM-2000EX II microscope (JEOL Co., Tokyo, Japan) equipped with a side-mounted digital camera (Advanced Microscopy Technologies) operated at 60 kV, as described by us [34]. Hippocampal tissues from 5xFAD and Non-Tg mice were fixed in a solution containing 4% paraformaldehyde and 2% glutaraldehyde in 0.13 M sodium cacodylate buffer (pH 7.2). After several washes in the same buffer, samples were post-fixed with 1% osmium tetroxide in 0.13 M sodium cacodylate buffer for 2 h, rinsed sequentially in buffer and distilled water, and then dehydrated through a graded ethanol series. Dehydrated tissues were infiltrated overnight at room temperature with a 1:1 mixture of Embed-812 resin and acetone, followed by three incubations in 100% Embed-812 resin for 2 h each. Samples were embedded in resin and polymerized at 65 °C overnight. Ultrathin sections (60–65 nm) were cut using a Leica EM UC7 ultramicrotome and mounted on 200-mesh copper grids. Sections were stained with Uranyless and lead citrate before imaging at 60 kV. For quantitative analysis of structural disorders using IPR-based molecular-specific mass density fluctuations, a total of 10-12 micrographs of mitochondria from each mouse were collected. A total of 4-5 images per mouse were examined in each group (Non-Tg and 5xFAD).

*2.2.5. IPR Quantification using TEM Images*

Mitochondria are called the powerhouse of the cell as they produce most of the energy required for cells/neurons. We wanted to study mitochondrial subcellular structural changes in AD. The IPR analysis is described in the sections above. A similar analysis was used to study mitochondrial structural disorder during the progression of AD. For this analysis, TEM images of Non-Tg and 5xFAD mice brain tissues from the hippocampus region were processed in ImageJ to isolate mitochondria, with only the mitochondria and background regions turned black, as shown in Figure 1. In the IPR analysis, we considered only IPR values arising from mitochondria. Interestingly, both the Mean and the standard deviation of IPR values increased with disease progression.

For TEM imaging, the refractive index/optical properties are indirectly related to the TEM images through mass density. The TEM intensity is proportional to the charge density, which in turn is proportional to the mass density. As it has been discussed in detail [14]:

$$\frac{dI_{TEM}}{I_{TEM}} \alpha \frac{d\rho}{\rho} \alpha \frac{dn}{n} \qquad (7)$$

Based on this, the TEM intensity can be converted to the tissue's refractive index, and the same IPR formalism can be applied.

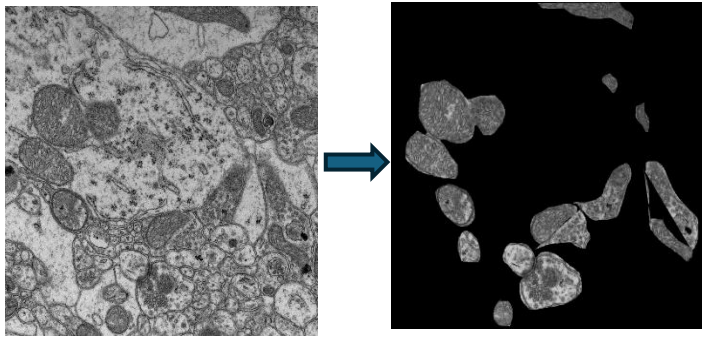

**Figure 1.** Transmission electron microscopy (TEM) image of mitochondria from the 5xFAD mouse model (left). The image was processed with ImageJ to isolate individual mitochondria (center), and the background was further separated for analysis (right).

*2.3. Novel object recognition*

The novel object recognition assay was employed to evaluate recognition memory, a component of learning and memory, based on the natural inclination of mice to investigate novel stimuli [35,36]. The test was conducted in an open-field arena (40 cm × 40 cm × 30 cm) containing two objects as described by us [34]. During the habituation phase (Day 1), Non-Tg and 5xFAD mice were individually allowed to explore the empty arena freely. To minimize olfactory interference, the chamber was cleaned with 70% ethanol between trials to eliminate olfactory cues. On Day 2, animals were initially presented with two identical, familiar objects (red outlined) positioned along adjacent walls. After a 2-hour retention interval following this acquisition phase, one of the familiar objects was replaced with a novel object (white outline), while the other remained unchanged. Each trial lasted 5 minutes, and all sessions were video recorded and analyzed using EthoVision XT software (Noldus, Wageningen, The Netherlands) for automated tracking of object exploration time. The time spent interacting with the novel versus the familiar object was recorded, with increased exploration of the novel object interpreted as an indication of intact recognition memory.

## 2.4. Immunofluorescence Staining

Immunofluorescent labeling was carried out following previously established protocols by our group [15,37,38] . In brief, mice were anesthetized using isoflurane, and brains were rapidly extracted, post-fixed in 4% paraformaldehyde, and cryoprotected in 30% sucrose prepared in 0.1 M phosphate-buffered saline (PBS) at 4 °C for 48-72 hours. Coronal brain sections (25 $\mu m$ thick) were prepared using a cryostat. Sections were rinsed in PBS, then blocked in a solution containing 5% bovine serum albumin (BSA; Sigma-Aldrich) and 0.3% Triton X-100. Subsequently, sections were incubated overnight at 4 °C with the primary antibodies ionized calcium-binding adaptor molecule 1 (Iba-1; 1:500, Synaptic System) and the 4G8 anti-Aβ antibody (BioLegend). After primary antibody incubation, sections were treated with fluorescent secondary antibodies, either Alexa Fluor 488 anti-chicken or Alexa Fluor 555 -anti-mouse (1:500, Invitrogen). Following washes, sections were counterstained with DAPI (Vector Laboratories) and coverslipped. Imaging was performed with a fluorescence microscope at 50× and 200× magnification, using 5× and 20× objectives, respectively. Image acquisition parameters (laser intensity, gain, and magnification) were kept consistent across all experimental groups.

## 2.5. Quantification for Mitochondrial DNA Copy Number

Mitochondrial DNA (mtDNA) copy number was quantified to assess mitochondrial abundance in hippocampal tissue of 5xFAD and Non-Tg mice. Total DNA was isolated using a commercial extraction kit (Bio-Rad), ensuring high-quality nucleic acid suitable for downstream amplification. Quantitative PCR (qPCR) was performed using primer sets specific for mtDNA and nuclear DNA (nDNA) sequences. The ratio of mtDNA to nDNA amplification served as an index of relative mtDNA copy number. The primer sequences used in this analysis are provided in the results section below.

## 3. Results
### 3.1. PWS Analysis of Mice Brain Tissues
#### 3.1.1. PWS Analysis of Cortical tissues

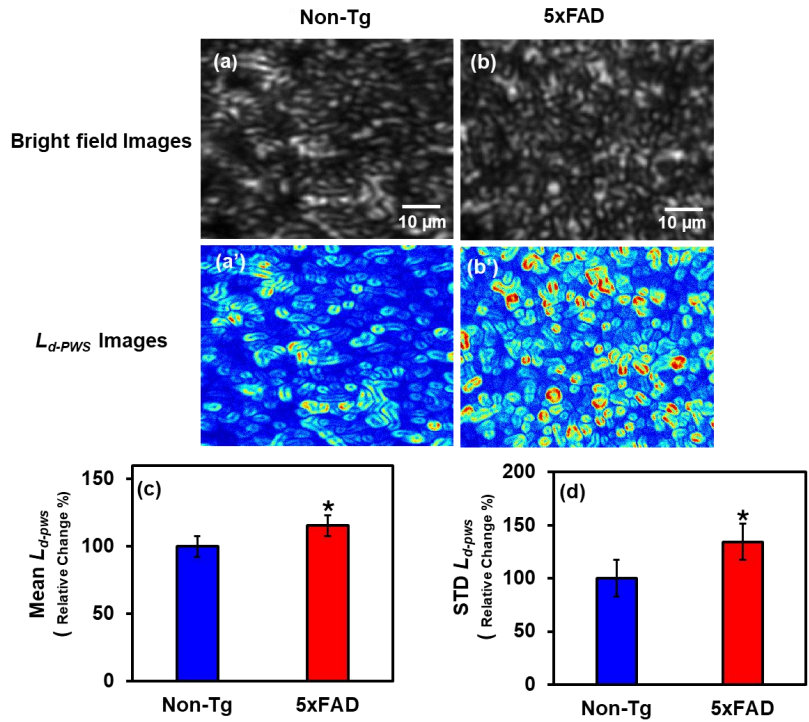

**Figure 2**. (a) and (b) are the bright field images of cortical brain tissue from 4-month-old Non-Tg and 5xFAD brain, respectively, whereas (a) and (b') are their respective $L_{d\text{-}PWS}$ images. (c) and (d) are the bar graphs of the relative Mean and STD *of $L_{d\text{-}PWS}$ in* Non-Tg and 5xFAD mice tissues from the cortex region. There is a 15.5% increase in Mean $L_{d\text{-}PWS}$ in 5xFAD compared to Non-Tg mice. In STD $L_{d\text{-}PWS}$, there is an increase of 34.3% in STD $L_{d\text{-}PWS}$ in 5xFAD compared to Non-Tg. ( * Student's t-test P value < 0.05, n = 6-8 per category).

Using the methods described above, we used the PWS setup to analyze thin 10 μm mouse brain tissue from the cortex region, captured by the CCD in the visible spectrum. We determined the structural disorder strength $L_{d\text{-}PWS}$ at each image pixel. A 2D color map was constructed by calculating $L_{d\text{-}PWS}$ for each pixel, with blue indicating less disorder and red the most. Figure 2 (a) and (b) show the bright field images of the Non-transgenic (Non-Tg) and 5xFAD mice tissues, whereas (a') and (b') are the 2D $L_{d\text{-}PWS}$ color maps of those images. In color maps, there are more yellow and red spots in 5xFAD tissue than in Non-Tg tissue, indicating greater structural disorder, more refractive index fluctuations, and greater mass density changes. We also conducted a statistical analysis to compute the Mean and standard deviation (STD) of $L_{d\text{-}PWS}$ to compare Non-Tg and 5xFAD tissues. Figure 2 (c) and (d) show the Mean and STD bar graphs of $L_{d\text{-}PWS}$ comparing Non-Tg and 5xFAD mice brain tissues. There is a significant increase in Mean $L_{d\text{-}PWS}$ of 15.5% in 5xFAD tissue relative to Non-Tg, whereas in STD, there is an increase of 34.3% in STD $L_{d\text{-}PWS}$ in 5xFAD relative to Non-Tg. Hence, 5xFAD has a significant increase in structural disorder compared to Non-Tg, indicating greater mass density fluctuations.

*3.1.2.* PWS analysis of Hippocampus Tissues

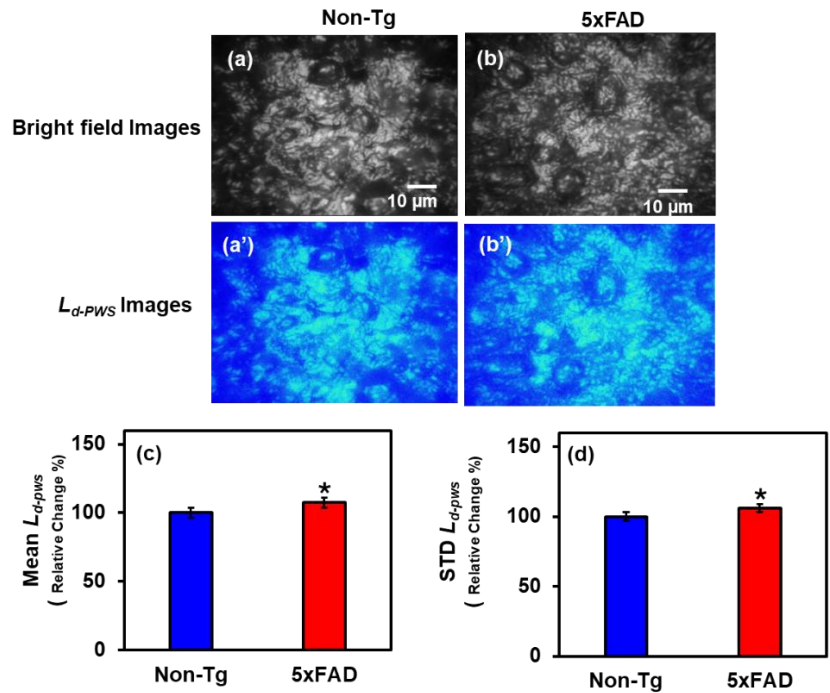

**Figure 3**. (a) and (b) are the bright field images of the Non-Tg and 5xFAD brain tissues of mice from the hippocampus region, whereas (a') and (b') are their $L_{d\text{-}PWS}$ images. (c) and (d) are the bar graphs of relative change in Mean and STD $L_{d\text{-}PWS}$ of Non-Tg and 5xFAD mice tissues from the hippocampus region of the brain. There is a 7.5% increase in Mean $L_{d\text{-}PWS}$ in 5xFAD compared to its control. In STD $L_{d\text{-}PWS}$, there is an increase of 6% in 5xFAD compared to Non-Tg. (* Student's t-test P value < 0.05, n= 6-8 per category)

Furthermore, we examined hippocampal tissue from 5xFAD and Non-Tg mouse brains to investigate structural changes in the hippocampus associated with AD. Figure 3 (a) and (b) show the bright field images of Non-Tg and 5xFAD mice hippocampal tissues, respectively, whereas (a') and (b') are the respective 2D color maps of $L_{d\text{-}PWS}$ of those images. The statistical comparisons are shown in the bar graphs in (c) and (d), which depict the Mean and STD of $L_{d\text{-}PWS}$ Non-Tg and 5xFAD. There is an increase of 7.5% and 6% in the $L_{d\text{-}PWS}$ values in 5xFAD brain tissues compared to Non-Tg, respectively. This increase in structural disorder strength indicates greater fluctuations in tissue mass density.

## 3.2. IPR Analysis of Mice Brain Tissues

### 3.2.1. IPR Analysis of Cortex Region

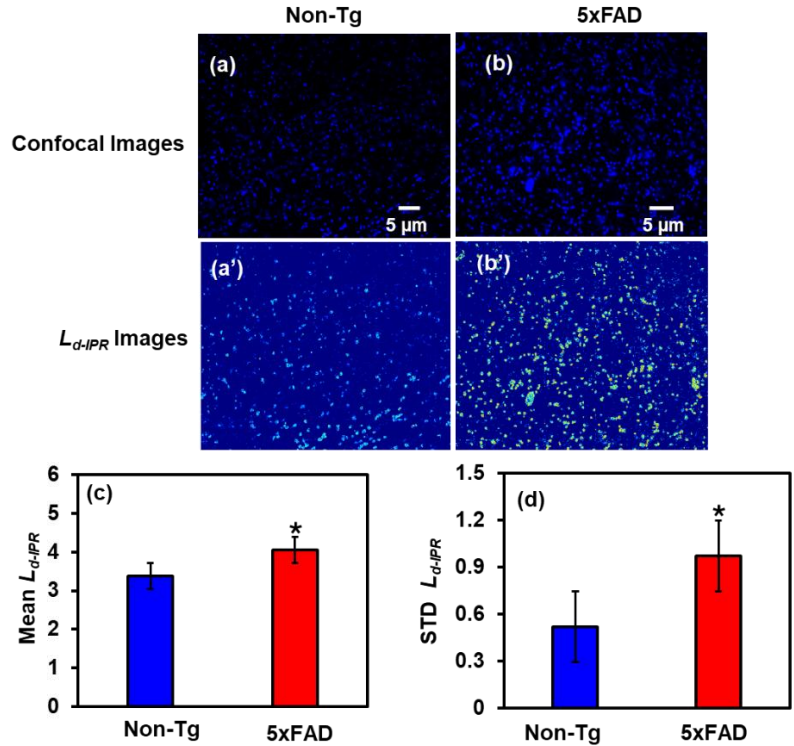

**Figure 4**. DNA/Chromatin molecular-specific structural disorder by confocal-IPR images and $L_{d\text{-}IPR}$ (a) represents the confocal image of Non-Tg mouse brain tissue stained with DAPI; (b) represents the 5xFAD mice brain tissue from the cortex region, DAPI-stained targeting DNA/chromatin. (a') and (b') are the corresponding images of $L_{d\text{-}IPR}$ for Non-Tg and 5xFAD DAPI-stained brain tissues, respectively. Bar graph for Mean and STD $L_{d\text{-}IPR}$: (c) illustrates the average $L_{d\text{-}IPR}$ values for DAPI-Stained DNA/Chromatin Non-Tg and 5xFAD mice brain tissue(cortex). (d) shows the STD $L_{d\text{-}IPR}$ values for Non-Tg and 5xFAD of the DAPI-stained tissue. The average and STD $L_{d\text{-}IPR}$ for 5xFAD are relatively higher than their respective control. Mean $L_{d\text{-}IPR}$ is 20% and STD $L_{d\text{-}IPR}$ is remarkably 82% higher in the 5xFAD tissues compared to Non-Tg. This is due to structural disorder in DNA/chromatin, leading to cell/tissue damage and abnormalities, and altering macromolecular arrangements. (* Student t-test P-value < 0.05, n = 6-8 per category)

After investigating tissue-level structural changes, we perform the molecular-specific IPR technique to examine nanoscale structural alterations in the nuclear region of mouse cortical brain cells, targeting DNA/chromatin. We used DAPI-stained confocal imaging targeting nuclear DNA/chromatin. We took at least 5-7 confocal micrographs of each type to compute the IPR values. We calculated the Mean and STD of the $L_{d\text{-}IPR}$ after removing the background (non-DNA) region from the confocal micrographs. They were compared between Non-Tg and 5xFAD samples to identify differences in DNA structural disorders.

Figure 4 (a) and (b) are the confocal images of DAPI-stained Non-Tg and 5xFAD mice brain tissues(cortex), (a') and (b') are their corresponding IPR images. We quantified the results by plotting bar graphs of the Mean and STD $L_{d\text{-}IPR}$ values for Non-Tg and 5xFAD mouse brain tissues, shown in (c) and (d), respectively. There was a 20% increase in the Mean $L_{d\text{-}IPR}$ value in 5xFAD compared to Non-Tg, whereas the rise in STD $L_{d\text{-}IPR}$ was remarkably 82% in 5xFAD compared to its control. This increase reflects structural disorder within DNA/chromatin, leading to changes in the arrangement of macromolecules.

### 3.2.2. IPR Analysis of Hippocampus Region

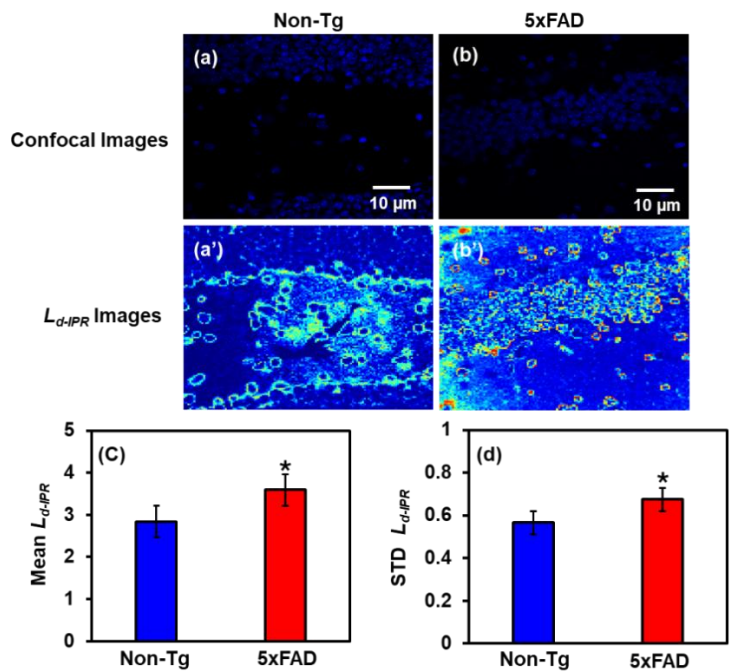

**Figure 5.** DNA/Chromatin molecular-specific structural disorder by confocal-IPR images and $L_{d-IPR}$ (a) represents the confocal image of Non-Tg mice brain tissue (hippocampus) stained with DAPI (b) represents the 5xFAD mice brain tissues (hippocampus) of DAPI-stained targeting the DNA/Chromatin. (a') and (b') are the corresponding images of $L_{d-IPR}$ for non-Tg and 5xFAD DAPI-stained brain tissues, respectively. (c) and (d) show the Mean and STD $L_{d-IPR}$ values for Non-Tg and 5xFAD of the DAPI-stained tissue. The average and STD $L_{d-IPR}$ for 5xFAD are relatively higher than their respective Non-Tg. Mean $L_{d-IPR}$ is 26% and STD $L_{d-IPR}$ is remarkably 19% higher in the 5xFAD tissues compared to Non-Tg. This is due to the structural disorder in DNA/Chromatin, leading to cell/tissue damage and abnormalities. (* Student's t-test P-value < 0.05, n = 6-8 per category)

The brain tissues from the hippocampus region of the mice were also studied by staining with DAPI, a nuclear stain. Figure 5 (a) and (b) are the representative DAPI-stained confocal images of Non-Tg and 5xFAD mice brain tissues, whereas (a') and (b') are their corresponding IPR images depicting the concentration of structural disorder in those tissues. To quantify the results, bar graphs were used to better understand alterations in local structural disorder in DNA. An increase of 26% in Mean $L_{d-IPR}$ in 5xFAD mice DNA compared to their control, as represented in (c). Similarly, in STD $L_{d-IPR}$ the fluctuation rose by 19% in 5xFAD mice compared to their Non-Tg controls, as shown in (d).

AD causes chromatin within the nucleus to misfold, altering local structural disorder. This increases local spatial heterogeneity, leading to greater fluctuations in mass density. An increase in $L_{d-IPR}$ can also be linked to DNA damage in brain cells caused by AD. In our previous publication, we verified the DNA damage analysis on brain cells due to AD in humans[15].

### 3.3. Changes in Mitochondria Structure in 5xFAD Mice: TEM Study

We selectively investigated cropped TEM images, focusing solely on mitochondrial structural alterations within hippocampal cells, as shown in Figure 6. (a)-(b) shows the cropped TEM images of mitochondria of Non-Tg and 5xFAD mice brain cells, whereas (a')-(b') are their corresponding IPR images. To quantify the results, a bar graph was plotted showing the Mean and STD $L_{d-IPR}$ for Non-Tg and 5xFAD. We found that the Mean mitochondrial IPR increased by 18% in 5xFAD mice compared with the control, whereas the STD IPR increased by 21% shown in (c)-(d). This indicates an increase in mitochondrial structural disorder, which can be linked to an increase in mitochondrial DNA (mtDNA) mutations/damages in AD, leading to mitochondrial fragmentation and ultimately neuronal death [39,40].

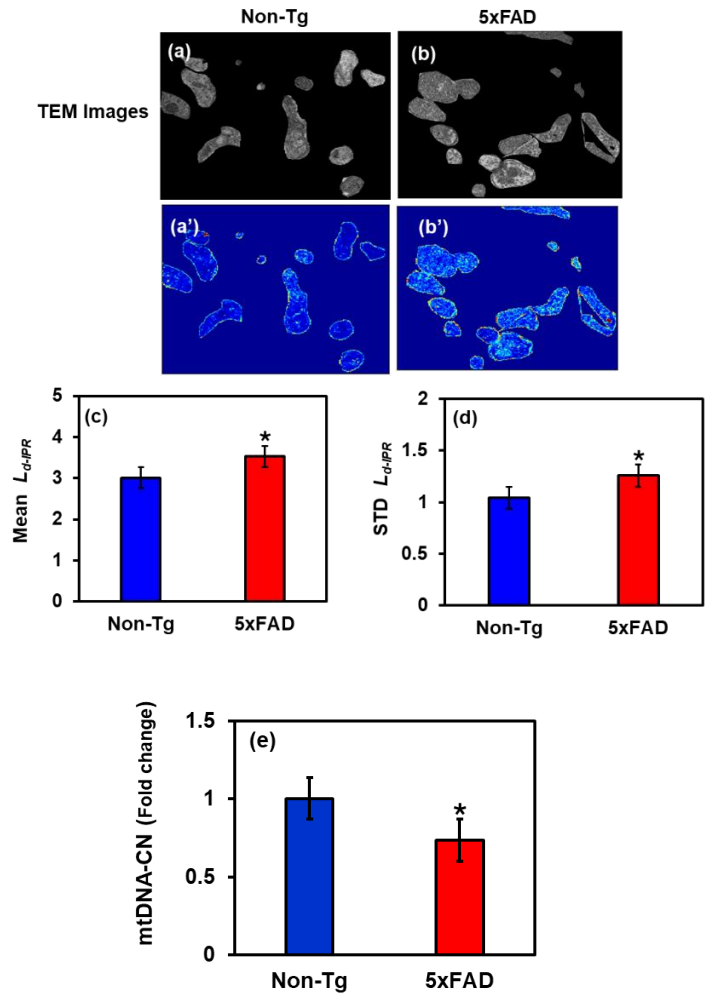

**Figure 6.** (a) and (b) are the cropped TEM images of mitochondria of Non-Tg and 5xFAD mice brain tissues from the hippocampus region. (a') and (b') are their corresponding $L_{d\text{-}IPR}$ images, respectively. Bar graph representation of Mean and STD $L_{d\text{-}IPR}$ values of mitochondria from Non-Tg and 5xFAD mice brain tissues. (c) shows there is an increase in Mean IPR values by 18% in AD than in the control, whereas (d) depicts an increase of 21% in STD $L_{d\text{-}IPR}$ values in AD than in its control. (e) represents relative mtDNA Copy number decreased in the hippocampus of 5xFAD mice by 27% compared to Non-Tg mice, showing impaired mitochondrial maintenance and early bioenergetic stress. (* Student's t-test P-value < 0.05, n= 10-12 per category)

## 3.4. Behavioral Study

To assess cognitive function, 5xFAD mice and Non-Tg controls were subjected to the Novel Object Recognition (NOR) test. The details of NOR are discussed in section 2.3. In the NOR assay,

5xFAD mice showed a significantly lower recognition index than their Non-Tg counterparts, suggesting deficits in recognition memory. Non-Tg mice spent more time interacting with the novel object than the familiar one, reflecting intact memory function. In contrast, 5xFAD mice failed to show a marked preference for the novel object, suggesting impaired discrimination. This diminished novelty preference implies disruption in memory encoding or retrieval mechanisms. Our results support the notion that Aβ-related pathology negatively affects brain regions critical for memory, such as the hippocampus and cortex.

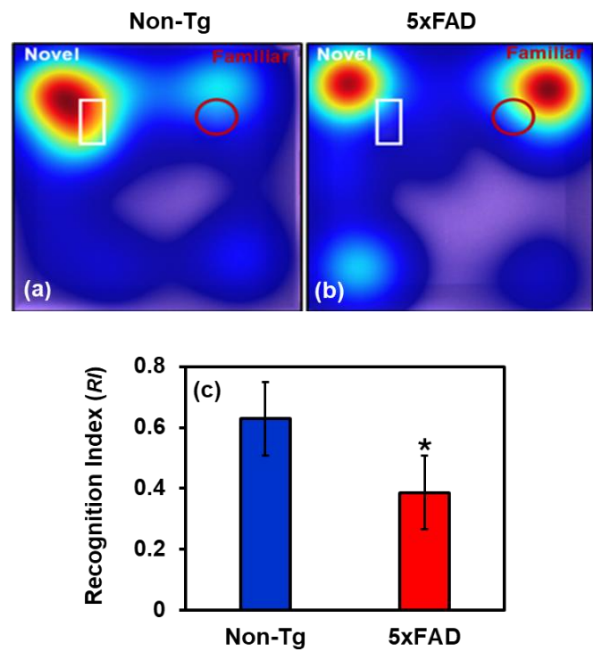

**Figure 7.** Spatial exploration patterns and recognition memory performance in 5xFAD and Non-Tg mice during the NOR test. Representative heat maps ((a)-(b)) display the distribution of time spent by mice across novel and familiar objects of the arena during the retention phase. The novel object was placed within the white-outlined region. The recognition index (RI), calculated as the proportion of time spent exploring the novel object relative to the total object exploration time, was significantly lower in 5xFAD mice, indicating impaired Non-spatial recognition memory. Quantitative analysis confirmed a notable reduction in RI in 7-month-old 5xFAD animals compared to Non-Tg controls, as shown in (c). Data are expressed as Mean ± SEM (n = 6 per group); (* Student's t-test P-value < 0.05)

*3.4.1.* Microglial Activation and Aβ Accumulation in the Brain of 5xFAD Mice

*3.4.2.*

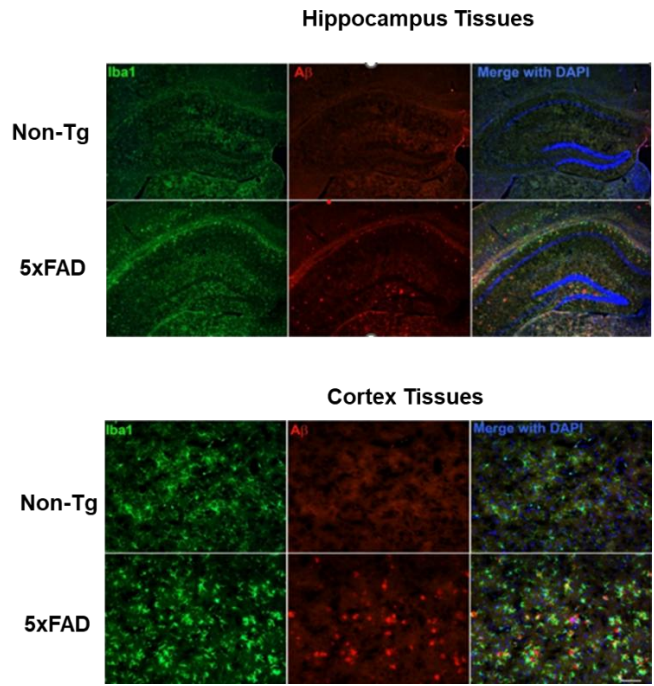

**Figure 8.** Microglial activation and amyloid-β (Aβ) accumulation in the hippocampus and cortex of 5xFAD and Non-Tg mice. Top panels: Representative immunofluorescence images of the hippocampal region stained for Aβ plaques (4G8, red) and microglia (Iba1, green) in 5xFAD mice and Non-Tg littermates, captured at 5× magnification. The lower panel shows representative cortical sections from the same groups, highlighting Aβ deposition (4G8, red) and microglial labeling (Iba1, green), imaged at 20× magnification. Images demonstrate increased plaque load and microglial activation in 5xFAD brains compared to Non-Tg controls. n = 3 mice/group

Immunofluorescence analysis revealed marked differences in Aβ plaque burden and microglial activation between 5xFAD mice and their Non-Tg littermates. In the hippocampus, 5xFAD mice exhibited extensive Aβ plaque deposition, visualized by 4G8 immunostaining, accompanied by pronounced microglial clustering, as indicated by increased Iba1-positive cells. These plaques were absent in Non-Tg mice, which also showed a sparse microglial profile, indicative of a homeostatic state. Similarly, cortical sections from 5xFAD mice showed a substantial accumulation of 4G8-positive Aβ plaques. In these regions, Iba1-positive microglia appeared hypertrophic and densely aggregated around the plaques, suggesting a reactive phenotype. In contrast, cortical tissues from Non-Tg controls exhibited no Aβ immunoreactivity and displayed microglia with morphology consistent with a non-activated state. These findings demonstrate significant amyloid pathology and reactive microgliosis in both hippocampal and cortical regions of 5xFAD mice, reflecting hallmark features of AD-like neuropathology. The spatial association between Aβ deposits and activated microglia further supports a potential role for neuroinflammation in disease.

*3.5.Mitochondrial DNA Analysis: Relative mtDNA*

Relative mtDNA copy number was significantly decreased in the hippocampus of 5xFAD mice compared to age-matched Non-Tg controls. This reduction in relative mtDNA indicates impaired mitochondrial maintenance and early bioenergetic stress associated with amyloid pathology, potentially heightening neuronal vulnerability in this Alzheimer's disease model. The statistics are depicted in the bar graph in Figure 6(e).

**Table 1:** Primers for mitochondrial DNA copy numbers in mouse

| Primer | Sequence (5′→3′) | Locus | | Species | Product (bp) |
| --- | --- | --- | --- | --- | --- |
| **mtDNA_mF1** | cagaaacaaaccgggccc | NC_005089.1 | 3322 - 3339 | Mouse | |
| **mtDNA_mR1** | gccggctgcgtattctac | NC_005089.1 | 3404 - 3387 | Mouse | 83 (with mtDNA_mF2) |
| **nDNA_mF1** | ccagggagagctagtatctagg | NC_000072 | 122150920 - 0941 | Mouse | |
| **nDNA_mR1** | ctggtcatgggagaaaaggc | NC_000072 | 122151095 - 1076 | Mouse | 176 (with nDNA_mF2) |

## 4. Discussion

In this study, we conducted a comprehensive, systematic investigation of the 5xFAD mouse model of AD, integrating dual-photon imaging, behavioral testing, and histopathological analyses. By combining dual-photonics techniques of PWS and IPR, we identified nanostructural alterations in the mass density of brain cells and tissues within the hippocampus and cortex. These nanoscale alterations closely paralleled Aβ plaque accumulation, microglia activation, and cognitive decline, underscoring a direct link between biophysical nanostructural disorder and AD-related neuropathology. Together, these findings provide a nanoscale perspective on the structural alterations underlying neuronal dysfunction in AD.

Quantitative analysis of PWS measurements revealed pronounced nanoscale structural alterations in 5xFAD mice compared with Non-Tg controls. The observed elevation in $L_{d\text{-}PWS}$ parameters in the 5xFAD mouse brain reflects the emergence of nanoscale structural disorder within cortical and hippocampal neurons, consistent with the progressive disorganization of the cellular microenvironment in AD. Such increases in optical heterogeneity likely stem from pathological protein generation, particularly Aβ and tau, which perturb local refractive index landscapes by altering membrane integrity, cytoskeletal stability, and extracellular matrix composition. The comparatively greater disorder in the cortex suggests a regional sensitivity to early amyloid pathology, in agreement with previous reports identifying cortical regions as initial sites of dense plaque deposition and neuroinflammatory activation [41–43]. Parallel alterations detected through IPR analysis further highlight the nuclear vulnerability underlying AD pathology. Elevated $L_{d\text{-}IPR}$ values in 5xFAD neurons reflect enhanced nanoscale mass-density fluctuations within chromatin and DNA domains, indicative of oxidative stress, DNA fragmentation, and disrupted chromatin compaction. Such nanoscale nuclear alterations are consistent with extensive DNA damage and chromatin disorganization reported in post-mortem human AD brains, where oxidative lesions, DNA double-strand breaks, and impaired DNA repair have been widely documented[37,44,45]. These changes represent early markers of genomic instability and transcriptional dysregulation, events that precede synaptic loss and neuronal dysfunction. Collectively, the convergence of PWS and IPR findings underscores that nanoscale structural disorder is not merely a secondary consequence of neurodegeneration but a fundamental aspect of the disease cascade. This emerging evidence suggests that biophysical disorganization at the chromatin and cellular levels may play a causal role in driving cognitive decline and the progression of AD pathology.

To further elucidate the subcellular mechanisms underlying nanoscale structural disorder in AD, we studied hippocampal mitochondria using TEM imaging combined with IPR-based quantitative analysis. Mitochondria from 5xFAD mice exhibited markedly elevated Mean and STDIPR values, indicating enhanced mass-density fluctuations and nanoscale disorganization relative to Non-Tg controls. These changes may reflect alterations in mitochondrial ultrastructure, including disruption, swelling, and fragmentation of cristae, morphological features commonly associated with impaired bioenergetics and oxidative damage. These nanoscale abnormalities likely arise from alterations in mitochondrial DNA (mtDNA), or overwhelming ROS loads, both of which compromise electron transport and ATP synthesis.

Mitochondrial dysfunction constitutes a critical nexus in AD pathogenesis, linking early nanoscale instability to downstream events such as energy failure, calcium imbalance, and apoptosis. The elevated IPR values observed here thus provide a quantitative nanoscale signature of mitochondrial distress that may precede overt neuronal degeneration. Reduced mtDNA copy number and impaired mitochondrial biogenesis, well documented in both human AD brains and transgenic models [46–48]. This further highlights the essential role of mitochondrial structural integrity in maintaining neuronal homeostasis. Because mtDNA encodes key components of the oxidative phosphorylation machinery, its loss directly diminishes ATP production while heightening oxidative stress [47,49]. In AD, these deficits exacerbate neuronal vulnerability by amplifying ROS production, disrupting calcium signaling, and activating cell death cascades [50,51]. Moreover, declining mtDNA copy number correlates with synaptic loss and cognitive impairment across aging and AD models [47,48,52,53], underscoring its central role in neurodegenerative progression. Taken together, our findings demonstrate that light-scattering–based nanoscale analyses, including $L_{d\text{-PWS}}$ and $IPR$, can sensitively detect mitochondrial and cellular structural abnormalities preceding histopathological lesions. Quantifying structural disorder across spatial scales provides a potential biophysical biomarker for early AD detection and offers mechanistic insights into how nanoscale disorganization of organelles contributes to neurodegeneration.

Behavioral analysis revealed that 5xFAD mice exhibited a significantly reduced recognition index compared with Non-Tg controls, consistent with pronounced cognitive impairment [54,55]. This behavioral deficit coincided with pronounced amyloid pathology and robust microglial activation in the hippocampus and cortex, regions that are fundamental to memory processing. These findings align with prior studies that highlight the 5xFAD model's fidelity in replicating key features of human amyloid pathology [55,56]. Histological analyses revealed a marked increase in Aβ plaques, indicative of extensive extracellular aggregation of Aβ peptides. These deposits were consistently encircled by clusters of Iba1-positive microglia exhibiting morphological features of activation, suggesting a phenotypic shift from a homeostatic to a pro-inflammatory state. This reactive microgliosis is emblematic of AD-associated neuroinflammation and is thought to contribute to synaptic impairment, neuronal dysfunction, and further amplification of Aβ pathology [57,58]. Activated microglia, through the release of pro-inflammatory cytokines and ROS, generate a neurotoxic milieu that exacerbates neurodegeneration [57,59]. Over time, this chronic inflammatory environment is likely to induce structural remodeling of brain tissue, disrupting local cytoarchitecture. Beyond the cellular and tissue level, sustained microglial activation and Aβ deposition may also cause alterations at the nuclear level. Increasing evidence suggests that nuclear architecture is sensitive to chronic oxidative stress and inflammatory cues[60,61]. In neurodegenerative settings, such stressors may initiate chromatin remodeling, histone modifications, and accumulation of genotoxic insults. Our previous work demonstrated elevated neuroinflammation and genotoxic stress in the brains of 5xFAD mice [37,62], supporting this notion. These alterations are reflected in changes detected by PWS and IPR analyses, which capture increased mass density fluctuations and chromatin reorganization at resolutions beyond conventional fluorescence microscopy. These nanoscale architectural changes may serve as early biophysical markers of pathology, as we previously reported in human AD brain tissue[15].

Notably, the cognitive deficits observed in 5xFAD mice may stem not only from classical macroscopic hallmarks such as Aβ plaques but also from more subtle nanoscale perturbations in nuclear and chromatin architecture. Together, our findings support the concept that AD pathology reflects a continuum of structural disruptions, ranging from extracellular protein aggregation and neuroinflammation to intranuclear architectural disorganization, that collectively impair cognitive function. These findings underscore the value of incorporating high-resolution biophysical approaches, including PWS or IPR imaging tools, into preclinical studies.

## 5. Conclusions

This study offers a multimodal, multiparameter view of AD in the 5xFAD mouse model by combining dual-photon imaging with histopathological, ultrastructural, and behavioral analyses. All the results are summarized in Figure9. Our findings reveal that nanoscale structural disorder arises early in AD and correlates with overt neuropathological changes and cognitive impairment. Elevated $L_{d-PWS}$ and $L_{d-IPR}$ values reflect increased mass density fluctuations within mitochondrial and nuclear compartments, suggesting that amyloid deposition, neuroinflammation, and oxidative stress collectively destabilize brain architecture across multiple spatial scales. These results demonstrate that optical biophysical approaches such as PWS and IPR can sensitively detect subcellular alterations that are invisible to conventional imaging techniques. Quantifying such nanoscale disorder provides a robust, powerful framework for understanding the structural underpinnings of neurodegeneration and identifying potential early biomarkers of AD and related dementia. A limitation of our study is the use of a single transgenic model and the lack of longitudinal in vivo validation. Expanding these approaches to additional AD models that constitute both amyloid and tau pathology will be critical for establishing a more comprehensive framework linking nanoscale structural disorder with behavioral and molecular outcomes across various stages of disease progression. This approach will advance the translational potential of nanoscale biophysical metrics as early biomarkers of AD and possibly other neurodegenerative disorders.

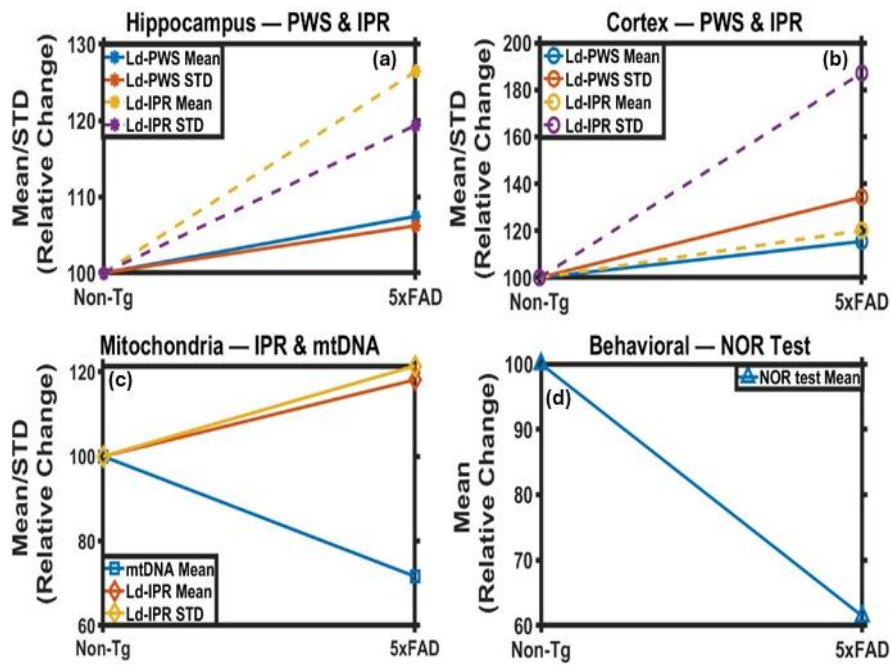

**Figure 9:** Summary Charts of changes in 5xFAD tissues with respect to Non-Tg : (a) represents $L_{d\text{-}PWS}$ and (b) represents $L_{d\text{-}IPR}$ results from the hippocampus and cortex tissue of the mouse brain, respectively for Non-Tg and 5xFAD mice. The Mean and STD of $L_{d\text{-}PWS}$ and $L_{d\text{-}IPR}$ values are increasing for hippocampus and cortex for 5xFAD vs. Non-Tg. (c) The Mean and STD of $L_{d\text{-}IPR(TEM)}$ values are also increasing, while mtDNA is decreasing for 5xFAD vs Non-Tg, for the hippocampus region. (d) Behavioral study results from the NOR analysis, the value decreases for 5xFAD wrt Non-Tg.

**Funding:** This work was partially supported by the National Institutes of Health under grant R21 CA260147 to PP. MMK's works on Alzheimer's disease and related dementia are supported by NIH grant R03AG075597, Alzheimer's Association Award AARG-NTF-22-972518, and Department of Defense Award Number HT9425-23-1-0043.

**Acknowledgments:** We thank the MS state and UTHSC imaging facilities for their imaging support.

**Conflicts of Interest:** The authors declare no conflicts of interest.